\documentclass{article}

\usepackage{arxiv}

\usepackage[utf8]{inputenc} 
\usepackage[T1]{fontenc}    
\usepackage{hyperref}       
\usepackage{url}            
\usepackage{booktabs}       
\usepackage{amsmath,amssymb,amsfonts}       
\usepackage{nicefrac}       
\usepackage{microtype}      
\usepackage{lipsum}		
\usepackage{graphicx}
\usepackage{xspace}
\usepackage{textcomp}

\newcommand{\solname}{BAD\xspace}
\newcommand{\longname}{Blockchain Anomaly Detection\xspace}

\newtheorem{definition}{Definition}

\title{\solname: a \longname solution}

\author{{\hspace{1mm}Matteo Signorini} \\
	Nokia Bell Labs\\
	France, Paris\\
	\texttt{matteo.signorini@nokia.com} \\
	\And
	{\hspace{1mm}Matteo Pontecorvi} \\
	Nokia Bell Labs\\
	France, Paris\\
	\texttt{matteo.pontecorvi@nokia.com} \\
	\And
	{\hspace{1mm}Wael Kanoun} \\
	Thales\\
	UAE, Abu Dhabi\\
	\texttt{wael.kanoun@thalesgroup.com} \\
	\And
	{\hspace{1mm}Roberto Di Pietro} \\
	Hamad Bin Khalifa University\\
	Qatar, Doha\\
	\texttt{rdipietro@hbku.edu.qa} \\
}

\renewcommand{\undertitle}{Arxiv e-print}

\hypersetup{
pdftitle={\longname solution},
pdfauthor={Matteo Signorini},
pdfkeywords={Blockchain, Security, Anomaly},
}

\begin{document}
\maketitle

\begin{abstract}
Anomaly detection tools play a role of paramount importance in protecting networks and systems from unforeseen attacks, usually by automatically recognizing and filtering out anomalous activities. Over the years, different approaches have been designed, all focused on lowering the false positive rate. However, no proposal has addressed attacks specifically targeting blockchain-based systems. In this paper we present \solname:  \longname. This is the first  solution, to the best of our knowledge, that is tailored to detect anomalies in blockchain-based systems. \solname is a complete framework, relying  on  several components leveraging, at its core,  blockchain meta-data 
in order to collect potentially malicious activities. 

\solname~enjoys some unique features:
(i) it is distributed (thus avoiding any central point of failure);
(ii) it is tamper-proof (making it impossible for a malicious software to remove or to alter its own traces);
(iii) it is trusted (any behavioral data is collected and verified by the majority of the network); and,
(iv) it is private (avoiding any third party to collect/analyze/store sensitive information).
Our proposal is described in detail and validated via both experimental results and analysis, that highlight the quality and viability of our \longname~solution.
\end{abstract}


\section{Introduction}
\label{sec:introduction}

The \emph{Internet of Things} (IoT) digital revolution has brought a wide range of smart devices in the global market that are remotely accessible via Internet and able to communicate and cooperate with each other.
This opens great opportunities from an application and service point of view, but it also creates new security challenges as devices are easily accessible from Internet \cite{8771413}.

To address the above introduced typology of threat, \emph{intrusion detection systems} (IDS) have been developed in the past as tools aimed at strengthening the security of complex networks and systems via capturing, monitoring, and analyzing the peers' traffic or, more in general, their behavior \cite{DiPietro:2008:IDS:1394397}. These approaches, usually based on log analysis and data correlation, aim at building attack models and mitigation strategies on top of them. 
Existing IDS can be classified based on their approach into two classes: signature recognition or anomaly behavior \cite{Javitz}. On the one hand, the first class leverages databases where signatures of well-known attacks are matched. These databases are then used as a reference model to detect future occurrences of such attacks. Hence, this approach is not able to recognize new attacks whose signatures are still unknown. On the other hand, anomaly detection approaches build models of normal behavior and rise alerts for deviations from such baselines.
Thus, the goal of an \emph{anomaly detection system} (ADS) is to build the normal behavior model and then to challenge it with new/unknown behaviors in order to analyze how close they are to the reference model.

\textit{IDS} and \textit{ADS} proved their functionalities so far, especially when based on trusted third parties that are responsible to build reference models and to alert end-users or end-devices if an unexpected behavior has been detected. 
We can consider the classic case of anti-virus companies that build and manage threat databases, which are later used to identify known threats or to predict zero-day attacks. However, this approach does not work for truly distributed peer-to-peer communities that lack trusted anchors or centralized management, as in blockchain-based applications. 
Firstly designed as a support tool for Bitcoin \cite{Zohar_2015}, the blockchain technology allows untrusted peers within open (i.e. permission-less) communities to agree on the status of a shared database, without the necessity to access trusted third parties. The only assumption is that the majority of involved peers is honest and willing to keep the protocol up and running against malicious users. However, has shown in real life applications, attackers can \textit{eclipse} their victims (i.e. manipulate honest nodes access to the mainstream global blockchain), thus reducing the number of honest peers participating in the overall blockchain network. Eclipsing a node allows or simplifies several types of attacks as shown in \cite{Singh06eclipseattacks,7467362}.

{\bf Contributions}

In this paper we propose \solname:  a general solution that leverages the features of blockchain to provide and Anomaly Detection Service. As an instance of its effectiveness, \solname 
allows the peers in a blockchain network to be protected against eclipse attacks by sharing information on previous attacks (i.e. by re-distributing malicious forks to the whole peer-to-peer community). 
To the best of our knowledge, our approach is the first one that leverages forks on a global scale to detect and prevent local threats. The core idea behind \solname is to collect local attack logs in the form of (hashed) malicious transactions. These logs are generated by \solname from an attack sequence injected by an attacker on isolated victims, and they are later reused to prevent similar attacks on uncorrupted nodes. More precisely, the attack logs (usually discarded in standard blockchain applications) populate a threat database that allows other potential victims to be resilient against zero-day attacks already discovered. The proposed solution is detailed and implemented. Achieved results show it quality and viability, and pave the way for future research along the highlighted directions.

{\bf Roadmap}
This paper is organized as follows: in Section \ref{sec:backgroundandrelated} the blockchain background technology is introduced as well as related previous works on anomaly detection systems. Section \ref{sec:ThreatModel} describes our threat model. Sections \ref{sec:OurSolution} and \ref{sec:ExperimentalTests} introduce respectively our solution and the related experimental results. 
In Section \ref{sec:Overhead} we discuss the  overhead analysis of \solname as well as its theoretical complexity, while Section \ref{sec:Discussion} addresses issues and limitations of our system. Finally, Section \ref{sec:conclusion} concludes the paper and introduces future work.
\section{Technology Background and Related Work}
\label{sec:backgroundandrelated}

In the rest of this paper we adopt the same notation used in \cite{watten16} to describe blockchain and, in general, state machine replication protocols.
We will only consider permissionless blockchain technologies, where a race among peers is established for mining blocks and rising potential forks (see Bitcoin, Ethereum and Tether which are the cryptocurrencies with higher market capitalization\footnote{https://coinmarketcap.com/}). We give some concepts and definitions from \cite{watten16} below, followed by a general description of a blockchain protocol.

\noindent
An \emph{output} is a tuple consisting of an amount of bitcoins and a spending condition. 
The latter is usually a valid signature associated with the private key of the spender address, however it can be generally a script which could be exploited by an attacker. \\
\noindent
An \emph{input} is a tuple consisting of a reference to a previously created output and arguments for the spending condition. This allows the transaction creator to spend the referenced output. We call \emph{UTXO} the set of unspent transaction outputs.

\begin{definition} \cite{watten16}
A \emph{transaction} is a data structure that describes the transfer of bitcoins from spender to recipients. The transaction consists of a number of inputs and new outputs. The former result in the referenced output spent (removed from the UTXO), and the latter being added to the UTXO.
\end{definition}

\begin{definition} \cite{watten16}
	A \emph{block} consists of a transactions' list, a reference to the previous block and a nonce. Each block contains those transactions that the block creator (called the miner) has accepted in its memory-pool since the previous block.
\end{definition}

\subsection{Blockchain Technology}
\label{sec:background}

Blockchain technologies are specifically designed to avoid single point of failures, i.e. those scenarios in which a single fault (either malicious or not) 
can affect the entire system by disrupting the provided service. 
These technologies  solve this problem by replicating the server nodes and orchestrating their interaction with clients thus, achieving fault-tolerant services.
As such,  the fundamental property achieved by blockchain technologies is the state machine replication (SMR), which is defined as follows (we will use the Bitcoin's terminology for the sake of simplicity):

\begin{definition} \cite{watten16}
	A set of miners achieves \emph{state replication}, if all the miners execute a (potentially infinite) sequence of transactions $t_1,t_2,t_3, \ldots$, in the same order. 
\end{definition}

State replication  is crucial to enforce the exact same state for all miners over time, while a set of transactions (issued by several users/wallets) is received and executed. Note that, miners are usually located on different machines to ensure that their eventual failures are independent. 
Although different in several aspects such as performances, permissions, provable security and computational completeness, any blockchain implementation satisfies the above definition.
As an example, a central blockchain's tool that differs among implementations is the \emph{consensus}  algorithm \cite{Fischer}. It solves the following problem, which is crucial in designing an efficient SMR protocol.
In the consensus problem, we consider a finite set of processes (or nodes in the network) $p_1,p_2, \ldots, p_n$ which communicate by exchanging messages.
These processes could fail and we will consider the worst case: the byzantine failure.
Initially, each process $p_i$ is in an \emph{undecided} state and proposes a value $v_i$ by broadcasting it to every other node. At the end, each node $p_i$ will decide the value of its \emph{decision variable} $d_i$. We can now formally define  consensus  as follows \cite{Coulouris}:

\begin{definition}
	A set of $n$ processes $p_1,p_2, \ldots, p_n$ achieves \emph{consensus} if the following properties hold:
	\begin{itemize}
		\item Agreement: the decision values of all the correct processes are the same;
		\item Integrity\footnote{In the literature, integrity is also called ``validity''.}: if the correct processes all proposed the same value $v$, then any correct process has set its decision variable to $v$;
		\item Termination: eventually each correct process sets its decision variable. 
	\end{itemize}
\end{definition}

\noindent   
In the remaining part of this section we give a brief review of how standard ADS systems work and provide an overview on how  we can build an ADS on top of the meta-data leveraged by a blockchain running a proof-of-work like \cite{10.1145/3396374} consensus protocol that generates local meta-data discarded at the time of block creation. We refer the reader to \cite{watten16} for a formal and more complete treatment of blockchain's protocols.

\subsection{Anomaly Detection Systems}
\label{sec:related}

By recognizing and then discarding, sanitizing, or otherwise nullifying outliers input that might exploit security vulnerabilities, ADS often play a central role in many computer security systems \cite{10.1145/1541880.1541882}. Formally, an ADS can be defined as a couple $(M,D)$, where $M$ is the reference model describing the expected behavior while $D$ is a similarity measure which specifies the actual behavior's deviation from $M$. 
Over the years, several ADS approaches have been proposed.

In statistical methods for anomaly detection, the system observes subjects' activities and generates different profile baselines to represent their behavior. 
Haystack was one of the earliest examples of statistical based ADS \cite{Smaha} which used a range of values that were considered normal and used to detect intrusions.  Machine learning based prediction tools can be used to guess the next expected values; thus, they can be used in ADS to build the reference model by predicting normal incoming events, given the current ones. 
It is then possible to detect anomalies by selecting those next events which are not the ones anticipated by the prediction tools \cite{Halavati_2007,Cintra_2007,Edward_Hinojosa_Cardenas_2012}.
Machine learning approaches study algorithms that allow systems to derive general behaviors from data, and which can be either supervised or unsupervised.
The first model is created from known clean data while the second is constantly analyzing data and modifying the behavior model without owning a previous one.
For example, Spectrogram \cite{conf/ndss/SongKS09} is a machine learning based statistical ADS for defense against web-layer code-injection attacks orchestrated by a network situated sensor that dynamically assembles packets to reconstruct content flows, and learn to recognize legitimate web-layer script inputs.
Taint-based techniques have been analyzed in ADS to avoid the false positives common issue. However, their applicability is limited by the need for accurate policies on the use of tainted data.
Cavallaro et al. \cite{Cavallaro2011-AnomalyDetection} developed a solution capable of detecting attack types that have been problematic for taint-based techniques, while significantly cutting down the false positive rate.

A preliminary report on the work in progress on \solname was published in \cite{8495798}. In those two pages we just exposed the general idea. In this contribution, we experimentally prove its viability and formally define the related framework. Note also that  \solname served  as a baseline for filing  a Nokia Bell Labs patent \cite{signorini2019blockchain}---a clear sign of its innovative and viable approach, poised to have a concrete impact on both industry and research.

\subsection{ADS Challenges}

ADS usually need to protect the reference model used to detect known and unknown threats \cite{Murtaza_2014,Kim_2015}. In host-based ADS (H-ADS) this database is stored locally while in network-based ADS (N-ADS) it can be either centralized on a trusted third party or distributed among the peers. 
	
The problem of having centralized data-storage and management systems which are susceptible to breaches becomes even worse in truly distributed networks such as the ones leveraging blockchain technologies \cite{Xu2016}. 
Furthermore, although a blockchain technology prevents several types of unexpected behaviors from malicious or compromised peers on a global scale, it does not eliminate attacks on a local scale. Indeed, local malfunctioning of the blockchain (see Section \ref{sec:ThreatModel}) are discarded and cannot be used by others to recognize attack sequences that get reused over time by an attacker. As a result, ADS tools aimed at protecting blockchain-based systems cannot solely rely on information appearing within the mainstream chain but also need to take into account local contexts, and share such information on a global scale.

Table \ref{tab:ads-related} lists some of the latest approaches in designing ADS systems on top of blockchain technologies \cite{9099189}. We have grouped these approaches in Table \ref{tab:ads-related} by highlighting (on the columns) four key properties as follow:

\begin{itemize}
	\item Approach: describes whether the solution uses blockchain technologies to: i) build a \textit{framework} for detecting anomalies; ii) as a simple (yet reliable) \textit{storage system} to keep track of ADS data built by other tools; and, iii) as \textit{other ADS approaches} that applied various techniques on top of blockchain meta-data; 
	
	\item Attack: identifies the vector through which malicious data is introduced within the system, i.e. how the adversary tries to subvert the system or control sensitive data. It can either be on-chain or off-chain. The former identifies attacks using the blockchain data structure to inject malicious code, while the latter identifies those attacks which are carried outside the blockchain;
	
	\item Data usage (for short DU): this is a boolean flag that identifies those solutions that leverage blockchain meta-data, usually discarded by the p2p network, to better understand and identify anomalies within the system;
	
	\item Data creation (for short DC): unlike the above property that leverages blockchain data to analyze anomalies, this boolean property identifies those solutions enriching the standard blockchain meta-data with additional information that could help other nodes in identifying anomalies.
\end{itemize}

\begin{table}[ht]  
	\caption{Related works on blockchain-based anomaly detection systems.} 
	\centering 
	\begin{tabular}{l c c c c} 
		\hline
		\hline  
		Solution & Approach & Attack& Data Usage & Data Creation  
		\\ [0.5ex]  
		\hline   
		S. Iyer \cite{8944586}& storage& off-chain& $\centerdot$ & $\centerdot$ \\
	    S. Sayadi \cite{8766765}& storage& on-chain& $\centerdot$ & $\centerdot$ \\
	    S. Morishima \cite{9029110}& other& off-chain& $\centerdot$ & $\centerdot$ \\
	    M. Salimitari \cite{9013824}& framework& off-chain& $\centerdot$& $\centerdot$ \\
	    Z. Il-Agure \cite{9075114}& other& off-chain& \checkmark & $\centerdot$ \\
	    X. Wang \cite{8883087}& framework& on-chain& $\checkmark$& $\centerdot$ \\
	    BAD (Our solution) & framework& on-chain& \checkmark& \checkmark\\
		\hline 
	\end{tabular}  
	\label{tab:ads-related}
\end{table}

As shown in Table \ref{tab:ads-related}, although there are other works focusing on the study of ADS applied to the blockchain technology, to the best of our knowledge, \solname is the first approach that designes an ADS framework which not only works with the blockchain meta-data (forks being created and discarded over time) but also enriches it by sharing on a global scale all those information that are typically  generated and stored on a local scale.
\section{Threat Model}
\label{sec:ThreatModel}

The solution proposed in this paper has been designed to be resilient against any class of attacks where a malicious entity can append its own transactions within the blockchain to inject malicious code in the system. However, for the sake of simplicity and clarity, we will use the well-known \emph{eclipse attack} \cite{Singh06eclipseattacks,7467362} to provide an example of these attacks, and how our solution counters them.

\begin{definition}
	A \emph{fake transaction} is a blockchain transaction used as a side channel to deliver an unexpected message.
\end{definition}

\begin{definition}
	A \emph{malicious transaction} is a special type of fake transaction in which the hidden message has the main purpose of attacking one or more peers within the network. 
\end{definition}

\begin{definition}
	A \emph{fake block} is a blockchain block that contains one or more fake/malicious transactions. Fake blocks can be either eventually discarded or accepted as part of the mainstream chain.
\end{definition}

The standard blockchain network (used in Bitcoin) has been designed to be decentralized and independent of any public key infrastructure. Indeed, each node connects to 8 other nodes stored in a list that is obtained by querying DNS seeders. In an eclipse attack, the attacker infects a node's list of IP addresses, thus forcing the victim's node to connect to IP addresses controlled by the attacker. Furthermore, the attacker also aims at filtering and manipulating victim's incoming connections. 

\begin{figure*}[!ht]
	\centering
	\includegraphics[keepaspectratio, width=0.8\linewidth]{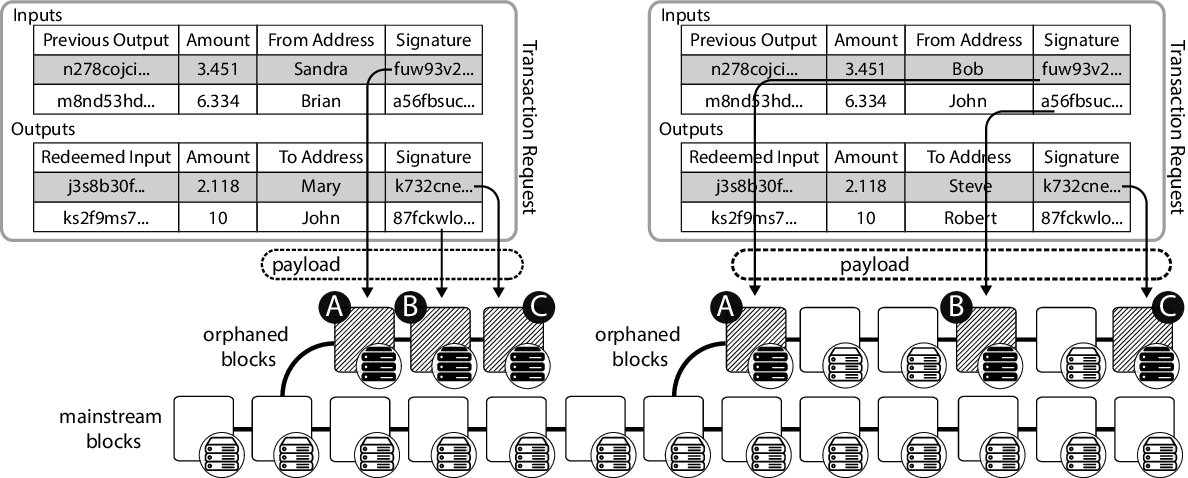}
	\caption{Example of malicious code embedded within orphaned blocks}
	\label{fig:EmbeddedMaliciousCode}
\end{figure*}

One way to execute an eclipse attack, is to repeatedly and rapidly forming unsolicited incoming connections to the victim by attacker's controlled IP addresses and then to wait until the victim restarts \cite{Heilman}. Hence, one challenge for the attacker is to control enough number of IP addresses in order to increase the probability that all the victim's outgoing connections will be directed to IP addresses controlled by him (see Section \ref{sec:ExperimentalTests}). Once the attacker has monopolized all the victim's connections, he can filter incoming blocks and send his own \emph{fake blocks} containing either \emph{malicious transactions} as it has been done in ZombieCoin \cite{Ali2015} (see Fig. \ref{fig:EmbeddedMaliciousCode}). For the above attack to succeed, we assume the following attacker's capabilities:

\begin{itemize}
	\item \textbf{Network Control}: the attacker can manipulate victims' connections in order to control their inbound and outbound traffic, thus being able to isolate them. This is a standard requirement for the eclipse attack;
	
	\item \textbf{Blockchain Control}: the attacker is capable of creating fake blocks which are sent to the victim. Their content is forged ad hoc by the attacker and usually contains a malicious payload.   
\end{itemize}

\paragraph{Liveness of the system.} As described in Section \ref{sec:TestResults}, we assume to have one or multiple powerful attackers who are able to perform eclipse attacks on one or several victims. However, they have to complete in a finite time window. This means that we always assume that the victim(s) will eventually: i) recognize a fork, ii) synchronize with the mainstream blockchain technology and iii) share all the information collected during the eclipse attack with other peers in the network.
\section{\solname: a \longname solution}
\label{sec:OurSolution}

The core idea behind \longname (\solname) consists in providing a new decentralized system based on the blockchain technology which leverages all the information collected from past forks. In blockchain-based applications, forks become more important as the chances to create their evolution for malicious purposes get higher. The rationale behind this approach is that while attacks may happen only once within a single device, when they are repeated over time against other devices they usually keep behaving in the same way. Hence, by collecting information on previous attacks, it could be possible to black list them and to prevent them within those peers that have not been attacked yet. In the following we first report the rationale that inspired \solname and provide and example of its applicability, and later discuss the complete application stack of our solution.

\subsection{\solname: rationale and example}	
\begin{figure}[ht]
	\centering
	\includegraphics[keepaspectratio, width=0.6\linewidth]{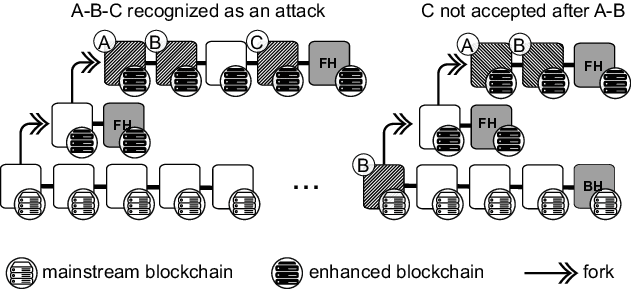}
	\caption{An overview of \solname being used as a tool to avoid known blockchain-based attacks to be repeated over time}
	\label{fig:IDSOverview}
\end{figure}

In our solution, information regarding chain forks and their orphaned blocks, is discarded (as usually done in classical approaches). Indeed we collect, enrich and share such information with other peers in the network. Shared information contains: i) the time at which the fork has started; ii) the time at which the fork has been detected; and, iii) the number and type of malicious transactions, if any, that has been identified within the fork. 
Fig. \ref{fig:IDSOverview} shows a toy example of how we build our enhanced blockchain. The longer chain in the figure represents the mainstream chain (eventually agreed by all peers) with \textit{block head} (BH) being the last blocks accepted. Shorter branches represent forks that happened in the past with \textit{fork head} (FH) being the last blocks accepted before a new fork was created. Last, but not least, the figure also contains an example of malicious payload being spread through the blockchain. Such payload is composed by three transactions labeled \{\textit{A}, \textit{B}, \textit{C}\} which, as explained in Section \ref{sec:ThreatModel}, can be either fake transactions or valid transaction embedding malicious code.
	
The collection of all fork-related information and the building of an enhanced blockchain made us able to design \solname as an ADS for blockchain-based applications. In fact, by having the enhanced blockchain agreed by all peers we only had to re-define $(M,D)$ (see Section \ref{sec:related}) to model our ADS. Indeed in our solution, \textit{M} is represented by the mainstream blockchain, thus describing the expected behavior, while \textit{D}(s) is represented by the fork(s), thus describing similarity measures and their deviation from \textit{M}. 
It is then possible to learn that, as shown in Fig. \ref{fig:IDSOverview}, \textit{A-B-C} have been previously labeled as an attack thus to prevent them from being re-executed on other peers.

Note that our solution is particularly efficient when the attacker, or the payload being spread, replicates the same operations (i.e. the transaction content) against every peer. For a more general approach, an additional ML/AI layer would be needed where a sequence of suspicious transactions can be compared with a set of malicious sequences (collected over time) in order to be identified as an attack and eventually prevented (see Section \ref{sec:conclusion}).  

\subsection{Application Stack}
\label{sec:architecture}

The standard blockchain application stack is structured in three layers: shared data, shared protocol and application.

\textbf{Shared Data Layer}: contains the core blockchain and its overlay network. It is still based on the core blockchain protocol but it is used to build networks (called sidechains \cite{8861821,
8946384,8835275}) that work in parallel to the mainstream chain to perform tasks that the mainstream chain cannot solve while still relaying on the same data structures. Whatever forms these overlay networks take, they all share the connection to the mainstream chain. Such a connection is used to bootstrap their own alternative solution by leveraging the mainstream peer-to-peer network;
	
\textbf{Shared Protocol Layer}: thanks to the blockchain it is now possible to develop decentralized applications with built-in data (transaction payload), validation processes, and transactions that are not controlled by any single entity;
	
\textbf{Application Layer}: applications built on top of the shared data layer and the shared protocol layer work very similarly to the ones we have nowadays. However, they inherit security, privacy and decentralization properties from the underlying blockchain technology. Hence, peers using these applications will be able to talk with each other and finally reach an agreement which is trusted even though no central authority has been used.

\begin{figure}[ht]
	\centering
	\includegraphics[keepaspectratio, width=0.5\linewidth]{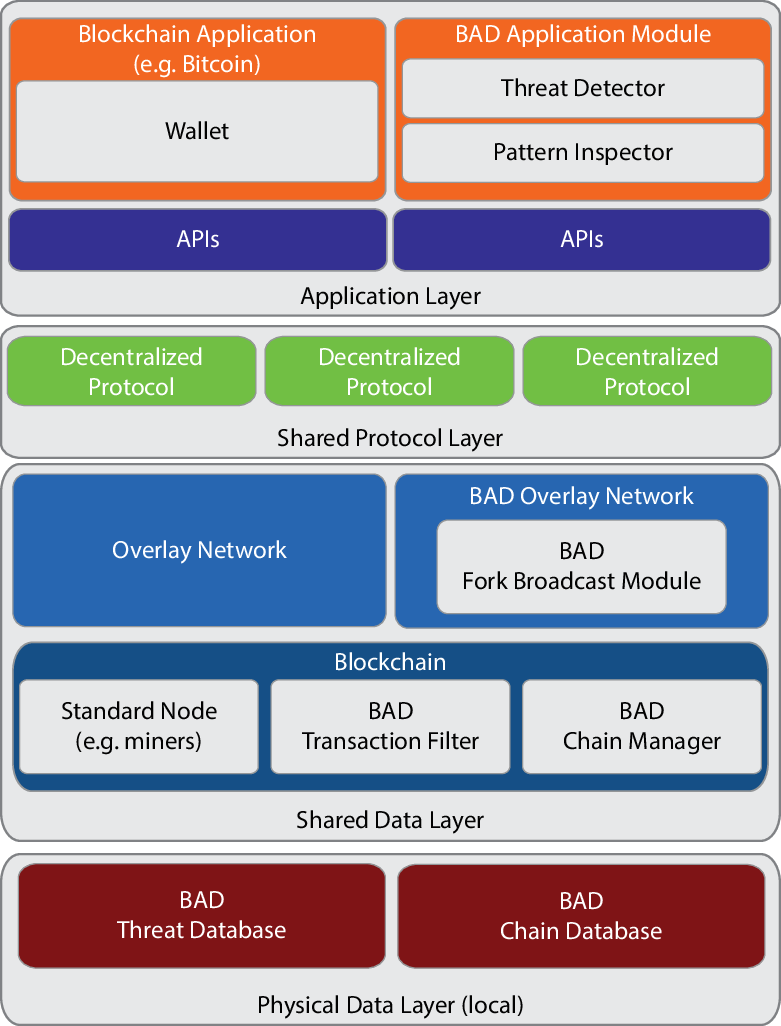}
	\caption{\solname application stack}
	\label{fig:BADStack}
\end{figure}

As shown in Fig. \ref{fig:BADStack}, \solname has been designed to be an ad hoc solution (i.e. a blockchain based application plug-in or a third party service) rather than being embedded within Bitcoin or any other specific blockchain application. The reason for such approach is that \solname does not rely on a specific blockchain and can be instructed to detect attacks on any blockchain application. Indeed, the core Bitcoin elements such as the \textit{wallet} and the \textit{miner} do not contain \solname elements but just interact with them. Here, we describe each \solname's module and how it interacts with standard blockchain applications:

\begin{itemize}
	
\item Transaction Filter (Tx Filter): intercepts standard blockchain messages and forward them to both the \textit{miner} and the \textit{chain manager}, thus not interrupting the standard protocol. Furthermore, it allows the collection of transactions meta-data;

\item Chain Manager: it is responsible to build our enhanced blockchain which, among the other elements, contains information on all forks that have been generated so far. It receives messages from the \textit{transaction filter} and retrieves additional missing information from the \textit{chain database} which finally stores our \textit{enhanced blockchains}. Last but not least, the chain manager notifies the pattern inspector if the enhanced blockchain has been updated and some threat analysis has to be applied;

\item Pattern Inspector: leverages the \textit{chain database} to detect unexpected behaviors. The inspection on the forks can be done with any approach ranging from signatures to heuristic static analysis and it is aimed at finding sequences of transactions which were found to be dangerous in the past;

\item Threat Detector: starting from the anomalies found by the \textit{pattern inspector} this module performs root-cause analysis by exploiting past blockchain activities (past blocks and transactions within them) to roll back all the operations done by the victim. Afterwards, all the attack information are collected within a \textit{threat database} which contains the information on all malicious patterns within the blockchain that have to be considered malicious (depending on the security policy being adopted).

\end{itemize}

\begin{figure}[ht]
	\centering
	\includegraphics[keepaspectratio, width=0.6\columnwidth]{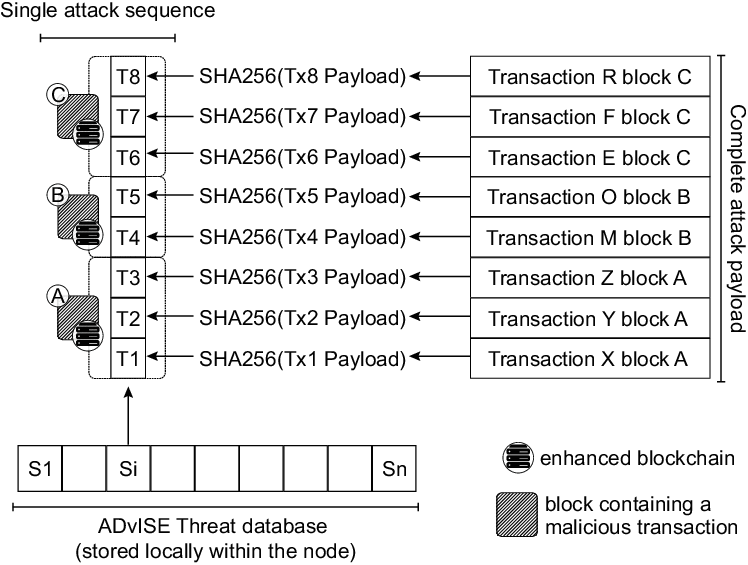}
	\caption{Implementation of the threat database in \solname}
	\label{fig:ThreatDatabse-Implementation}
\end{figure}

Fig. \ref{fig:ThreatDatabse-Implementation} shows a simple implementation of \solname's threat database. Here, recalling the toy example given in Fig. \ref{fig:IDSOverview} in which \{\textit{A}, \textit{B}, \textit{C}\} were found to represent chunks of a malicious payload, we show how this information is collected and later shared with other peers. \solname's threat database is basically a dynamic (i.e. not sized) array of array in which $S_{i}$ represents the i-th attack sequence detected while $T_{i}$ represents the hash of the i-th transaction which was found to contain part of the payload's attack sequence.

Information used to fill the threat database is provided by the pattern inspector and used by the transaction filter to avoid the repeating of known attacks. The filtering process is accomplished by the \solname's transaction filter module each time a new block is received and its overhead has been analyzed in Section \ref{sec:complexity}.

\section{Experimental Test}
\label{sec:ExperimentalTests}

In this section we show how \solname has been used in our experimental platform to prevent attacks across different networks thanks to the information collected from forks. 
The goal of this experiment was to detect forks on a given peer, that were caused by an eclipse attack, and then to share this information with other peers in order to build a reference model, aimed at detecting future occurrences of the same attack.

\subsection{Testbed}
\label{sec:TestResults}

\begin{figure}[ht]
	\centering
	\includegraphics[keepaspectratio, width=0.5\columnwidth]{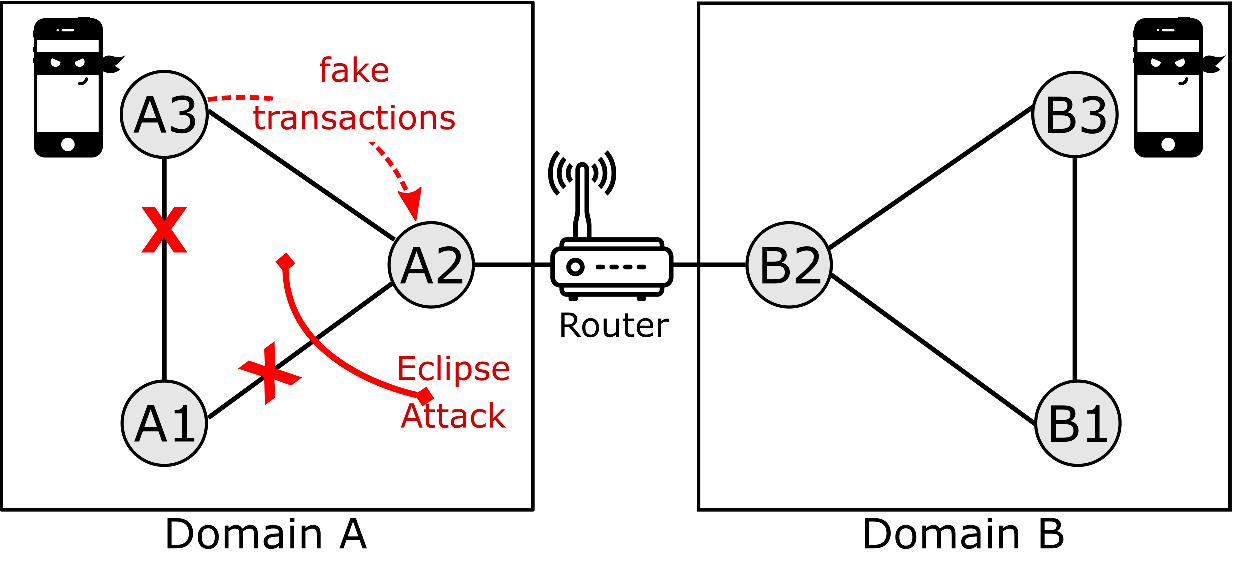}
	\caption{Eclipse Attack in Domain A}
	\label{fig:EeclipseAttack}
\end{figure}

\noindent
For simplicity, the testbed shown in Figure \ref{fig:EeclipseAttack} is only composed by two domains, A and B, that represent two separated private IP networks with a router between them. In domain A (\textit{B respectively}), we have deployed two full nodes and one lightweight client as follow: 

\begin{itemize}
	\item Full Nodes: two active full nodes\footnote{https://bitcoin.org/en/full-node\#what-is-a-full-node} \textit{A1} (\textit{B1} in domain B) and \textit{A3} (\textit{B3} in domain B) are deployed on a virtual machine with 4 GB of RAM with Linux Ubuntu 16.04 as a guest operating system. Both have been executed in regtest experimental mode\footnote{https://bitcoin.org/en/developer-examples\#regtest-mode}, i.e. a mode in which local testing environment can be created with instantaneous on-demand block generations and digital assets creation, without any real value. During the experimental tests, \textit{A3} in domain A (\textit{B3} in domain B) is assumed to be controlled by a malicious user;
	
	\item Client Node: as a lightweight client we used a Bitcoin Java BitcoinJ wallet (version 0.14.3)\footnote{https://bitcoinj.github.io} running on a 4GB RAM PC with Windows 8.1 installed as a guest operating system. This wallet acts as the victim of the eclipse attack and is labeled as \textit{A2} (\textit{B2} in B).
\end{itemize}

\textit{A1} (\textit{B1} in domain B) and \textit{A3} (\textit{B3} in domain B) are connected to each other, which means they can exchange blocks and agree on the longest chain---to do so we used on each node the following command:\\ \texttt{bitcoin-cli -regtest addnode IPaddr add}. \\Nodes in domains A and B are initially synchronized on the same blockchain---this blockchain is generated using the command: \texttt{bitcoin-cli -regtest generate X}, \\
that is meant to initialize $X$ blocks in the blockchain.

\subsection{Attack Detection and Prevention}

Based on the above testbed, we have implemented a real attack using bitcoin-cli commands. The attack aims at eclipsing a victim node and force it to accept some malicious blocks containing a payload. The attack, as well as the creation of our enhanced blockchain, has been implemented as follows:

\begin{enumerate}
	\item eclipsing \textit{A2} and forcing it to only communicate with \textit{A3} which is controlled by the attacker;
	\item stop \textit{A3} from exchanging blocks with \textit{A1} to avoid being detected by other nodes in the same domain. This has been implemented via executing: \texttt{bitcoin-cli -regtest addnode IPaddressofA1 remove} within \textit{A3};   
	\item make \textit{A3} sending to \textit{A2} three new blocks containing forged transactions. We have implemented this via the command: \texttt{bitcoin-cli -regtest generate 3};
	\item wait for \textit{A2} to send the above fake blocks as connected to the previous blockchain header and representing the longest chain received so far. Assuming that the above three new blocks, created by \textit{A3}, contains a malicious payload, we can conclude that \textit{A2} is compromised at this step;
	\item as the attack is completed, the eclipse on the victim is removed. Hence, \textit{A2} starts again to communicate with other peers in the same domain, and to receive blocks from them which eventually forces \textit{A2} to receive a longer chain that does not contain the above three fake blocks. At this point, and by leveraging on our \solname modules, \textit{A2} is capable of keeping track of the malicious blocks received and to share this information broadcasting it to the other peers in all domains.  
\end{enumerate}

 As a second phase we executed the same steps described above but this time within domain B. By leveraging \solname, and the information gathered so far from domain A, peers in domain B were able to detect and to prevent the attack from succeed. Indeed, we witnessed the (attempted) attack in domain B to behave as follows:
 
\begin{enumerate}
	\item \textit{B2} is eclipsed by forcing it to only connect to \textit{B3}, here controlled by a malicious user;
	\item \textit{B3} generates three malicious blocks which contain, among the others, the same three malicious transactions used in the attack against \textit{A2};
	\item  unlike \textit{A3}, \textit{B3} has now the knowledge of some malicious blocks/transactions that resulted in another peer being compromised. Indeed, as also shown in Fig. \ref{fig:IDSOverview}, \solname is able to detect blocks that are different but contains the same transactions (or a subset), in the same order, as previously received by \textit{A2}.
\end{enumerate}
 
The final result is the prevention of the complete attack as only a small subset of the malicious transactions is accepted (in our example accepted by \textit{B2}) before \solname recognizes them as malicious. As done by \textit{A2}, also \textit{B2} will share the information with other peers once it realizes that it was previously mining and elaborating on blocks that belonged to a malicious fork.
\section{Overhead Analysis}
\label{sec:Overhead}
The core elements introduced by \solname on the classical Bitcoin protocol are the broadcast of brand new forks, their orphaned blocks, as well as the detection of malicious transactions on new received blocks. In this section, we analyze the introduced bandwidth overhead to show that our solution is scalable and thus deployable within the standard Bitcoin network. In particular, the results of our analysis show that our system has minimal bandwidth consumption in comparison with the one consumed by standard nodes.

\subsection{Bandwidth overhead}
We have analyzed the overhead introduced by our solution in the worst-case scenario, i.e. the whole global Bitcoin fork activity to affect one single node named \textit{NX}. Our overhead is then defined as the amount of bandwidth that \textit{NX} consumes due to the fork broadcast introduced in \solname. To this aim, and to be rooted on real data, we have considered the maximum number of orphaned blocks discarded by the Bitcoin community during last year. We are interested in the total number of orphaned blocks because it includes those used to attack the victims (see Section \ref{sec:ThreatModel}). Furthermore, we assume this number to have a small variance since a smart adversary, to stay hidden in the network, would not create an anomalous number of orphaned blocks. A more abstract, and less constrained, analysis is given in Section \ref{sec:complexity}. 

To analyze \solname's overhead, we have designed the p2p network surrounding our \textit{NX} node. By construction, nodes in the Bitcoin network create a random graph with randomness emerging from the selection of outgoing connections. In the vanilla Bitcoin protocol, each node attempts to keep a minimum of 8 outgoing connections at all time. However, it has been observed that, on average, a Bitcoin node has 32  outgoing connections \cite{6688704}. Furthermore, the total number of orphaned blocks discarded during 2016 \footnote{https://blockchain.info/charts/n-orphaned-blocks} was 141 with a maximum block size of 0.993201 MB. 
As such, in our worst-case scenario, we consider all those 141 orphaned blocks (of the maximum size) to be collected and re-distributed in broadcast by \textit{NX}. To broadcast all these blocks with their transactions, \textit{NX} would send broadcast messages to its neighbors, which sum up to the global size of $32 \times 0.993201 \times 141=4.481$ GB per year. 
It is important to highlight that the total number of orphaned blocks is independent of the node's bandwidth. Hence, our worst-case scenario can be applied to any node: from lightweight SVP clients to relay nodes or miners. Furthermore, the total node/month upload bandwidth could vary according to nodes capabilities and ISP resources: it could require an initial 150 GB/month of uploaded data (which is the minimum recommended upload data plan to run a Bitcoin core\footnote{ https://bitcoin.org/en/bitcoin-core/features/requirements}) and reach values up to 300 GB/month or more.

\begin{figure}[ht]
	\centering
	\includegraphics[keepaspectratio, width=0.5\columnwidth]{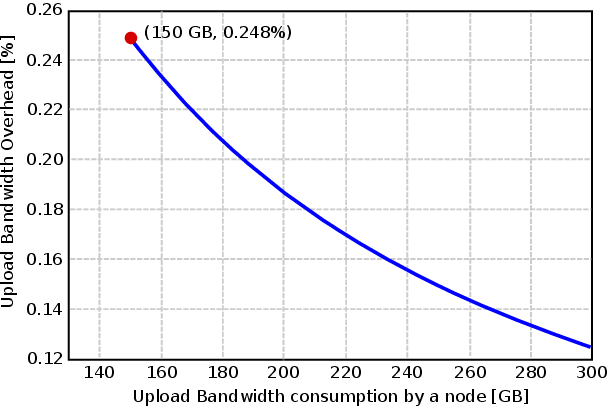}
	\caption{Overhead introduced by the system as a function of the bandwidth consumption of a node}
	\label{fig:OverHeadChart}
\end{figure}

\noindent
Fig. \ref{fig:OverHeadChart} plots the result of our \solname's overhead ($Ovh$) analysis which is approximated by the following formula: 

\vspace{0.05cm}

\begin{equation}
	Ovh = \frac{\textrm{\solname data broadcast (per year)}}{\textrm{total data exchanged (per year)}} = \frac{4.481}{m \times 12}
\end{equation}

\vspace{0.05cm}

\noindent
where m is the average bandwidth consumption of a node per month. Fig. \ref{fig:OverHeadChart} shows the maximum overhead introduced in the case of 150 GB of data upload consumption which is of 0.248\%. The results is an overhead on the bandwidth of only 0.248\%. This latter supports the fact that  \solname can be smoothly deployable in the standard Bitcoin network. 

\subsection{Complexity}
\label{sec:complexity}

In the previous section we studied \solname's overhead in the worst case, i.e. with an attacker using Bitcoin's forks to spread malicious code. However, statistics and real data used for such analysis refer to natural forks appeared over time in the network due to its delay. 

In thus section, we analyze a more general use case where the attacker creates as many blocks as needed (thus also generating more forks in the system). The result, as shown in the remaining of this section, is that \solname's bandwidth overhead, in the worst case, can only be proportional (up to a constant factor in real cases) to the size $k$ of our Threat Database $\mathcal{T}$.
Let $S_1, \ldots, S_k$ be the malicious transaction sequences of $k$ attacks detected and stored in  $\mathcal{T}$. Each malicious sequence $S_i$ has a length of $\ell_i$ transactions injected by the attacker to complete attack $i$. We call \emph{partial sequence} $(PS_i, j)$ a subsequence of $S_i$ starting from the first transaction and ending with the $j$-th transaction of $S_i$. Note that $(PS_i, \ell_i)$ represents the full attack $i$. 
For each attack $i$ we can have at most $\ell_i - 1$ distinct partial subsequences.
Each node in the network maintains a set $U$ of partial transactions.
Given that $H(t)$ is the hash of a transaction $t$, every time $t$ is analyzed by a node, \solname performs two actions:
\begin{enumerate}
	\item If there is a partial sequence $(PS_i, j) \in U$ such that $(PS_i, j) || H(t) =  (PS_i, j+1)$, we replace $(PS_i, j)$ with $(PS_i, j+1)$ in $U$. Here $||$ is the standard concatenation function.
	\item If $H(t)$ represents the first block of a sequence $S_i$, then we insert $(PS_i,1)$ into $U$.
\end{enumerate}   
Finally, \solname checks if there is a $(PS_i, \ell_i)$ in $U$ and, in that case, discards the transaction $t$.
While the correctness of this approach follows from the construction, the additional computational cost (per transaction) incurred by each node in the network can be derived.
Note that, in the worst case (which is when every transaction of every attack has the exact same hash),
every transaction will create a new partial sequence $(PS_i, 1), \, \forall i$, plus it will increase at most $\ell_i-1$ existing partial sequences in $U$ for each attack $i$. This translates in the following number of steps:
\[W(t) = k + \sum_{i=1}^k (\ell_i-1) = \sum_{i=1}^k \ell_i\]
Since (in a real scenario) each attack sequence is no longer than a constant $c$ of transactions, the total work $W(t)$ for a given transaction will be at most $c\cdot k = O(k)$ where $k = |\mathcal{T}|$.
In case the size of $\mathcal{T}$ grows very quickly, pruning techniques can be adopted to adjust its dimension. For example, old or infrequent attacks could be discarded in favor of newly discovered ones.
\section{Discussion}
\label{sec:Discussion}

The solution proposed in this work requires an attacker capable of \textit{pushing} fake blocks into his/her victim, i.e. to make the latter believing that some fake blocks received have been already accepted within the mainstream chain. In blockchain-based applications, this outcome can be achieved with a broad range of attacks spanning from owning 51\% of the whole peer-to-peer network, to leveraging the structure of the overlay network to eclipse the victim. Blockchain-based applications make large use of overlay networks \cite{1610546}, i.e. connections forming a graph upon which a distributed application is implemented, as they allow to deploy network functionalities without changing the underlying infrastructure.

As described in Section \ref{sec:ExperimentalTests}, the experimental tests provided to support our solution have been obtained by implementing an eclipse attack on our blockchain network. This required some bitcoin-cli commands that forced our victim node into adopting the malicious nodes as its peers, thus accepting all blocks received by them. The eclipse attack deployed on our network was quite easy to accomplish due to the limited size of our network. However, the state of art on distributed systems shows that a wide range of countermeasures and defense techniques can be adopted against such attacks. The solutions proposed by Castro et al. \cite{10.1145/844128.844156} based on constrained routing tables as well as the one proposed by Simgh et al. \cite{10.1145/1133572.1133613} based on neighbor anonymous auditing  are just some example describing how the eclipse attack can be prevented. Although this may suggest that the solution proposed in this work is limited since not easily deployable  in real networks, it should be highlighted that the above defense techniques against eclipse attacks, are used to either make some strong assumption on the network size/structure or to prevent optimizations like \textit{proximity neighbor selection} \cite{10.1145/863955.863998}: an important and widely used technique to improve overlay efficiency. Last but not least, the continuous development of new peer-to-peer protocols, mining algorithms and consensus schemes can make new blockchain  applications more exposed to eclipse attacks. In this latter case, \solname would be easily adoptable to counter such a threat, as well as to provide a customizable platform to counter further threats.

\section{Conclusion}
\label{sec:conclusion}

In this paper we proposed \solname: the first \longname solution. In particular, \solname allows to detect anomalous transactions and to prevent them from being further spread.
Indeed, while forks can naturally appear in the blockchain life cycle due to the network delay, they can also be artificially forged by attackers and used to spread malicious activities within the chain. \solname enables the prevention of repeated attack occurrences by collecting malicious and building a threat database which is distributed (thus avoiding any central point of failure), tamper-proof, trusted (any behavioral data is collected and verified by the majority of the network) and private. 

We  detailed  \solname, and provided an analysis of its overhead, as well as a prototype implementation, showing as an example its effectiveness in detecting the dreadful eclipse attack. The achieved results show the the quality and viability of our solution, that could also be a starting point for further investigation in this domain.



\bibliographystyle{unsrt}
\bibliography{main}  

\end{document}